\documentclass[namedreferences]{kluwer}    

\newdisplay{guess}{Conjecture}
\usepackage[dvips]{graphicx}

\begin{document}                                                                                   
\begin{article}
\begin{opening}         
\title{What can biologists say about galaxy evolution ?} 
\runningauthor{Fraix-Burnet, Choler, Douzery}
\runningtitle{Biology and galaxy evolution}

\author{Didier \surname{Fraix-Burnet}}  
\institute{Laboratoire d'Astrophysique de Grenoble, BP 53, F-38041 Grenoble cedex 9, France}
\author{Philippe \surname{Choler}} 
\institute{Laboratoire de Biologie des Populations d'Altitude, BP 53, F-38041 Grenoble cedex 9, France}
\author{Emmanuel \surname{Douzery}}
\institute{Institut des Sciences de l'\'Evolution de Montpellier, F-34095 Montpellier cedex 5, France}
\date{September 15, 2002}

\begin{abstract}
 It is possible to borrow from a topic of biology called 
        phylogenetic systematics, concepts and tools for a logical 
        and objective classification of galaxies. It 
        is based on observable properties of objects - characters - 
        either qualitative (like morphology) or quantitative (like 
        luminosity, mass or spectrum). Distance
        analysis can readily be performed using a method called 
        phenetics and based on characters. But the most promising 
        approach is cladistics. It makes use of characters that can 
        exist in at least two states, one being ancestral
        and the other one derived. Objects are gathered depending 
        on the derived states they share. We illustrate a first 
        application of this method to astrophysics, that we name 
        astrocladistics, with dwarf galaxies from the Local Group.
\end{abstract}

\end{opening}

   \section{Introduction}
  
  Biologists have a long experience in the classification of complex objects that are in evolution. It dates back to Aristote (around 100 A.D.) who attempted a first classification of living organisms. A lot of different systems have been proposed, but a first revolution came with Linn\'e (around 1730) and his binomial nomenclature that is still in use today! Its success is to be found in the fact that the names are unrelated to properties of objects. A few years later, Adanson (in 1763) proposed that an objective classification scheme can only be obtained by using all characters of an object. This idea has been incredibly successful and gave birth to what would be called nowadays phenetics or multivariate distance analysis. It was not until Darwin (around 1850) that evolution was found to be at the origin of the hierarchy and diversity of living organisms. It is then another century before Hennig (in 1950) devised a methodology for introducing the evolution in the classification process itself. This is called cladistics and has revolutionized somewhat the evolutionary classification of species (see an interesting discussion in Stewart  \citeyear{Stewart}).
  
  Systematics (science of classification of living organisms) distinguishes three categories of classification: 1) from apparent look (based on a few characters), 2) from global similarity (based on all characters), 3) from common history (based on the evolution of characters).  
Until the XVIIIth century, biologists classified living organisms according to the first category, the traditional way. It consists of choosing one or two characters (fruits, flowers, legs, wings) depending on very subjective a priori assumptions. It yielded quite a messy picture of Nature, with organisms not fitting in a given classification system and even incompabilities between different systems.
The second category is phenetics and benefitted a lot from the Linn\'e's binomial nomenclature. It is not clear who first proposed a hierarchical organization for living species, but Linn\'e clearly defined the two names as to correspond respectively to {\em genus} and {\em species}. This produced the well known evolutionary trees.
Nowadays, these trees are mainly a result of cladistics analysis, that corresponds to the last category of classification. It does not compare the values of characters, like in phenetics, but compare the states of evolution of each character. This information can be gained from observations (paleontology), from models or theories, or from comparison with other groups of objects. In a sense, cladograms (trees obtained from cladistics) are visualizations of the evolutionary schemes hypothesized for all characters. 

Since Hubble in the 1930s, galaxies have always been classified according to the traditional way (e.g. Roberts \& Haynes \citeyear{robertshaynes} and references therein). Nomenclature heavily follows physical properties (morphology, size, activity, luminosity at a given wavelength, radio loudness, etc...). But an overwhelming amount of new data compels us to consider galaxies as very complex and diverse objects, and their history as a quite complicated puzzle. We propose that a lot can be gained by paralleling the extragalactic context with biology. Some fundamental and fascinating questions can be raised: how do we describe a galaxy? what is the life of a galaxy? what is the history of galaxies considered as a species ? what about the environment?
  Following Linn\'e's approach, it seems to us that a new nomenclature will have to be coined for galaxies, based on a much more objective classification.
  
\section{Astrocladistics}

Multivariate distance analysis (second category of classification) can easily be used because it relies on tools and concepts familiar to astronomers. A very few attempts have been made (see references in Roberts \& Haynes \citeyear{robertshaynes}), but, to our knowledge, have not yielded a new classification. It has the power to determine the most discriminant characters for an objective classification. 

Cladistics for astrophysics (astrocladistics) is probably the most promising approach. Basically, the only requirement for cladistics to be applicable and useful is the presence of a hierarchical organization of the diversity due to evolution (see a discussion in Brower \citeyear{brower}). Because galaxies are made of many different evolving processes, because galaxies evolve strongly through interactions, diversity is increasing with time and very probably organizes itself in a hierarchical way. (Note that ``hierarchy'' is here taken in the whole complexity of a galaxy, not only in mass or size.) Astrocladistics is the tool that will help us finding the hierarchical classification of galaxies. 

Astronomers are particularly well equipped for cladistics. In principle, for each character, two states have to be defined: one will be said ``ancestral'' and the other one ``derived''. In practice, it is possible to define more than two states (8 with current softwares), hence hypothesizing an evolutionary scheme for each character. Processes occurring in galaxies are relatively well known and can be modeled sometimes with great details. Observations are providing more and more detailed data, at larger and larger redshifts. Hence, character coding in astrophysics does not seem to be too much of a problem.

\section{Application to the dwarf galaxies of the Local Group}

We are in the process of applying astrocladistics to a sample of dwarf galaxies of the Local Group. Data are taken from \inlinecite{mateo}. There are 36 galaxies and 28 characters (properties). Each character is binned in up to 8 states, and these states are related to each other by a linear transformation series. This hypothesis is obviously debatable and will have to be made more sophisticated in subsequent studies. Among the 28 characters, some will certainly appear to be irrelevant after a complete analysis is performed. This work, still in progress, will be published elsewhere. But to illustrate the power of astrocladistics, we present what kind of results and interpretations can be obtained.

\begin{figure}
\centerline{\includegraphics[width=8 cm]{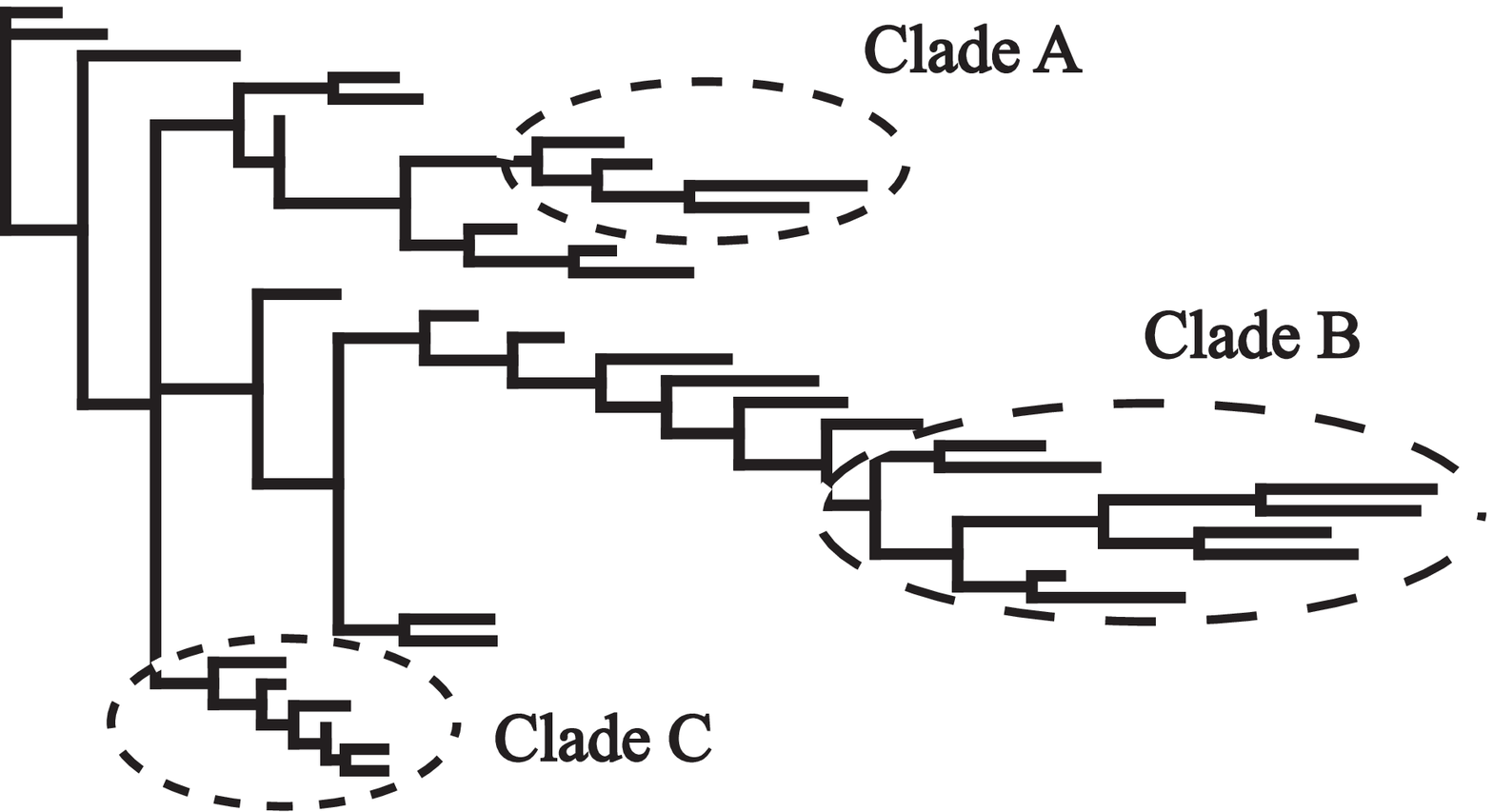}}
\caption{}
\label{tree}
\end{figure}

At the end of the analysis, groups (or clades) of galaxies are found on the cladogram (Fig.~1). This provides the classification. Galaxies in each group share a common history: it can be the same kind of past encounters, of environment during their lives, etc... In particular, one of the group might well be tidal dwarfs. To understand the history, it is necessary to look at which character changes on which branch of the cladogram. For instance, galaxies of the three clades defined on Fig.~1 have the following properties:

\begin{center}
	\begin{tabular}{ccc}
		\hline
   Clade A        &   Clade B        &   Clade C  \\
		\hline
	spheroidal      & irregular        & spheroidal \\
	low ellipticity & high ellipticity & high ellipticity \\
	low HI mass     & high HI mass     & low HI mass      \\
	low Fe/H        & all Fe/H         & high Fe/H        \\
	low M/L$_*$     & low M/L$_*$      & high M/L$_*$     \\
		\hline
	\end{tabular}
\end{center}

It is clear from the table that it is inappropriate to name these galaxies after their morphology or one other property, like birds cannot be described only by wings and mammals only by warm blood. A new taxonomy is to be invented, but this is not a trivial problem.




\end{article}

\begin{thebibliography}{}

\bibitem[\protect\citeauthoryear{Brower}{2000}]{brower} Brower, A.: 2000, `Evolution is not a necessary assumption of cladistics', {\it Cladistics} {\bf 16}, pp. 143-154
\bibitem[\protect\citeauthoryear{Roberts and Haynes}{1994}]{robertshaynes} Robert, M.S., Haynes, M.P.: 1994, `Physical parameters along the Hubble sequence', {\it ARA\&A} {\bf 32}, pp. 115-152
\bibitem[\protect\citeauthoryear{Stewart}{1993}]{Stewart} Stewart, C.-B.: 1993, `The powers and pitfalls of parsimony', {\it Nature} {\bf 361}, pp. 603-607
\bibitem[\protect\citeauthoryear{Mateo}{1998}]{mateo} Mateo, M.: 1998, `Dwarf galaxies of the Local Group', {\it ARA\&A} {\bf 36}, pp. 435-506

\end{thebibliography}
\end{document}